\documentclass[%
 reprint,
 notitlepage,
superscriptaddress,
tightenlines,
amsmath,amssymb,aps,
]{revtex4-1}

\usepackage[T1]{fontenc} 

\usepackage{esint}
\usepackage{graphicx}
\usepackage{textcomp}
\usepackage{amsmath}
\usepackage{amsfonts}
\usepackage{mathtools}
\usepackage{color}
\usepackage{mathrsfs}
\usepackage{empheq}
\usepackage{tcolorbox}
\usepackage{fancyhdr}
\usepackage{datetime}
\usepackage{lipsum}
\usepackage{graphicx}
\usepackage{dcolumn}
\usepackage{bm}
\usepackage{color}
\usepackage{xspace}
\usepackage{amsmath}
\usepackage{nccmath}
\usepackage{soul} 
\usepackage[normalem]{ulem} 
\usepackage{hyperref}

\newcommand*\bq{\mathbf{Q}}
\newcommand*\bQ{\mathbf{Q}}
\newcommand*\bu{\mathbf{u}}
\newcommand*\bn{\mathbf{n}}
\newcommand*\bp{\mathbf{p}}

\newcommand*\bI{\mathbf{I}}

\newcommand*\bE{\mathbf{E}}

\newcommand*\bx{\mathbf{x}}

\newcommand*\bOmega{\bm{\Omega}}

\newcommand{\fig}{Fig. }

\newcommand{\eqn}{Eqn. }
\newcommand{\eqns}{Eqns. }

\draft 

\begin{document}



\title{Mechanochemical Topological Defects in an Active Nematic}



\begin{abstract}
We propose a reaction-diffusion system that converts topological information of an active nematic into chemical signals. We show that a curvature-activated reaction dipole is sufficient for creating a system that dynamically senses topology by producing a concentration field possessing local extrema coinciding with $\pm\frac{1}{2}$ defects. The enabling term is analogous to polarization charge density seen in dielectric materials. We demonstrate the ability of this system to identify defects in both passive and active nematics. The model demonstrates that a relatively simple feedback scheme in the form of a PDE system is capable of producing chemical signals in response to inherently non-local structures in anisotropic media. We posit that such coarse-grained systems can help generate testable hypotheses for regulated processes in biological systems such as morphogenesis and motivate the creation of bioinspired materials that utilize dynamic coupling between nematic structure and biochemistry.
\end{abstract}

\author{Michael M. Norton}
\email{mmnorton@brandeis.edu}
\affiliation{Martin A. Fisher School of Physics, Brandeis University, Waltham, Massachusetts 02453, USA
}
\affiliation{School of Physics and Astronomy, Rochester Institute of Technology, Rochester, New York 14623, USA
}
\author{Piyush Grover}
\affiliation{Mechanical and Materials Engineering, University of Nebraska–Lincoln, Lincoln, Nebraska 68588, USA
}
\date{\today}
\maketitle

\section{Introduction}
\begin{figure}
\includegraphics[width=\columnwidth]{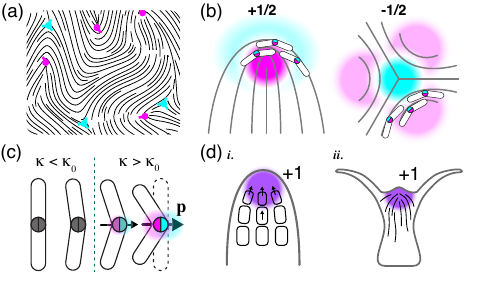}
\caption{(a) simulated active nematic laden with half-integer defects (+1/2 in magenta, -1/2 in cyan),  (b) schematic showing desired behavior of the defect sensing system: concentration extrema located at defect cores; (c) schematic of proposed enzyme that acquires a reaction dipole when deformed; the magenta side produces a product while the cyan side consumes it, (d) schematic of chemomechanical coupling in plant and animal morphogenesis, \emph{i}: the plant growth hormone auxin (purple) gradient driven by active transport in the direction of cell polarity (black arrows) and \emph{ii}: a hydra's actomyosin network (black lines) coupled to Wnt pathway activation (purple).}
\label{fig:schematic}
\end{figure}

Biological matter is defined by tightly integrated chemical and chemomechanical feedback loops that give rise to robust self-organized dynamics. These dynamical pathways enable directed motility, stimuli response, wound healing, and the reproduction of specific body plans. The field of active matter seeks to understand the broad principles underpinning these phenomena to advance biology and rationally design self-organizing materials. While active systems built from reconstituted cytoskeletal constituents and enzymes have yielded many insights into emergent dynamics, much of the regulatory machinery responsible for spatiotemporal patterning of activity is, by design, absent. Minimal models and experiments that couple form and activity are needed to explore self-regulating morphogenic matter. In control theory, closed-loop regulation consists of two key operations: \emph{sensing}, wherein the controller measures key features of the system to be controlled, and \emph{actuation}, the process by which the controller acts on this information to move a system towards a desired behavior. In this work, we focus on the sensing portion of the control process by developing a reaction-diffusion model that dynamically reports on structural features in an active nematic (\fig\ref{fig:schematic}a-c), its topological defects, by producing a concentration field.


Active nematics are a class of nonequilibrium materials composed of anisotropic rod-like constituents and characterized by their complex hydrodynamic flows driven by accompanying topological defects \cite{Marchetti2013,Decamp2015,Wagner2021,Alert2022}. Unlike passive liquid crystals, extensile nematics spontaneously create half-integer defects through the Simha-Ramaswamy bend instability \cite{aditisimha2002}. Controlling defect dynamics in microtubule-based nematics through imposed boundaries \cite{Norton2018a, Opathalage2019,Hardouin2019,hardouin2022,joshi2023,memarian2024,velez-ceron2024}, substrate friction \cite{guillamat2016, guillamat2017, thijssen2021}, external flows \cite{Rivas2020}, and externally patterned active stress \cite{Zhang2019, lemma2023, zarei2023} has been a fruitful area of research, but programmable autonomous dynamics are non-existent. Leveraging additional biochemical constituents to shape dynamics is a natural direction, but a theoretical framework for rationally choosing such components is needed.

\begin{figure*}
\includegraphics[width=\textwidth]{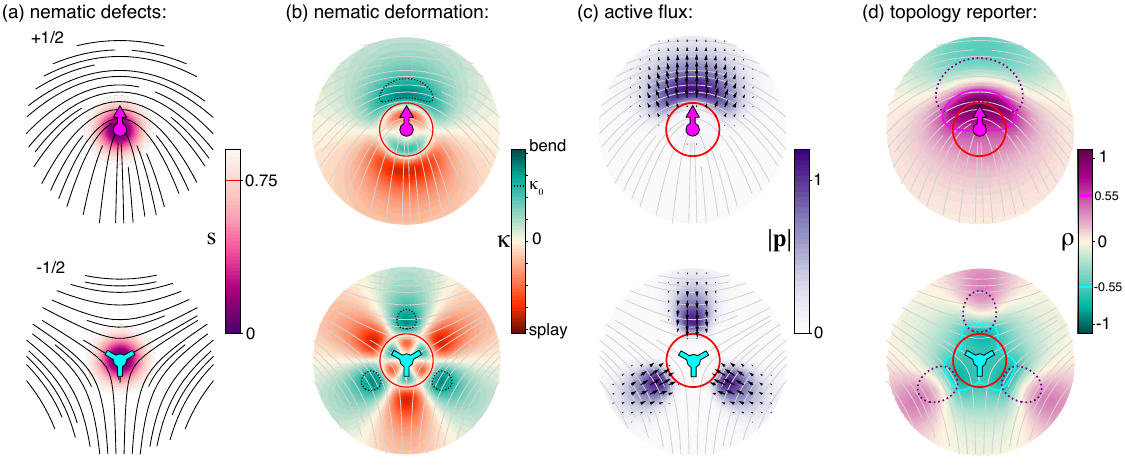}
\caption{Idealized behavior for isolated for $\pm\frac{1}{2}$ (top/bottom row) topological defects: panel (a) shows the director $\bQ$ and order $s$, (b) shows the bend-splay measure $\kappa$ (\eqn\ref{eqn:rhodynamics}) with the critical bend value, $\kappa_0$ that enables polar order, labeled (dashed black line), (c) shows the degree and direction of polar order $\bp$, and (d) shows the resulting concentration field $\rho$. Contours of $|\bp|=1$ (dashed purple) indicate the location of high dipole density and the contours of $\rho=\pm0.55$ (magenta, cyan) emphasize the asymmetry between $\rho$ interior and exterior to the defect cores for the two defect types.}
\label{fig:isolated}
\end{figure*} 

Biology provides abundant inspiration for linking topological features to chemistry in autonomous nematics. One example comes from plants wherein the coupling between polar order and chemical gradients organizes root growth. Throughout root tissue, the growth hormone auxin is transported from one side of cells to the other via membrane-bound PIN proteins \cite{leyser1999,benkova2003,heisler2010}. The orientation of the flux is established by an anisotropic distribution of these proteins along the cell, see arrows in \fig\ref{fig:schematic}d. Critically, the cells actively remodel their PIN distributions such that the flux orients towards high-auxin regions, reinforcing emerging gradients. These two processes together create a robust self-organizing dynamic that naturally co-localizes high auxin concentrations and +1 defects with the tips of roots, accelerating cell division at a specific spatial location thereby propagating the root. 

Another example of coupling between morphogenesis and topology comes from the animal kingdom. Defect arrangements in the hydra's supracellular actomyosin cable network are highly structured, locating its limbs, mouth, and foot, \fig\ref{fig:schematic}e. The organism cited by Turing as an exemplar of biological pattern formation has remained a model for morphogenesis because the locations of these defects are robust to destructive perturbations, permitting repeatable experiments \cite{Turing1952, Braun2018,Maroudas-Sacks2020,Maroudas-Sacks2021}. In the language of dynamical systems, hydra are able to regenerate their form from many initial conditions, suggesting that their body plan is a form of \emph{dynamical attractor} maintained by robust feedback mechanisms. For example, in ref. \cite{wang2020}, ectogenic Wnt patterning dictates the body axis while ref. \cite{ravichandran2024a} showed that soft compression of a hydra spheroid induces excess defects that eventually lead to multi-headed bodies. Taken together, there is growing evidence that structural anisotropy is not just a downstream outcome of biochemical processes, but an integral part of the biomechanical system regulating form. While recent models have explored the direct coupling of morphogen gradients to liquid crystalline order \cite{wang2023}, no models have considered the reciprocal process. Other theoretical and numerical studies have examined how active nematics possess an intrinsic ability to produce complex forms, but in the absence of any chemical dynamics \cite{hoffmann2022a, metselaar2019b, vafa2022}. 

Motivated by entwined topological and chemical processes in plant and animal morphogenesis and the desire to imbue model systems with similar attributes, we ask the question: can a simple mechanochemical reaction-diffusion system perform the geometry-to-chemical operations needed to \emph{sense} topological defects in a planar active nematic? Nematic defects are readily identified by eye or by algorithmically calculating the winding number of a region \cite{Chaikin2003,alexander2012}. However, both require a whole-picture view of the material because defects are inherently non-local structures. If we take the perspective of a material element embedded in the nematic, how can we know if we are near a defect?

We develop a bioinspired reaction-diffusion model that depends on local geometric features of the embedding nematic. \fig\ref{fig:schematic}b shows the idealized behavior. At the heart of our defect-sensing system is an active flux whose strength and orientation depend on the embedding nematic's structure. We envision this flux arising from a distribution of infinitesimal source-sink pairs, created by an orientable enzyme that catalyzes the production of a product on one end and the destruction of that product on the other, shown schematically in \fig\ref{fig:schematic}d. The concentration field created by this active process spontaneously highlights regions containing $\pm\frac{1}{2}$ defects.

\begin{figure*}
\includegraphics[width=\textwidth]{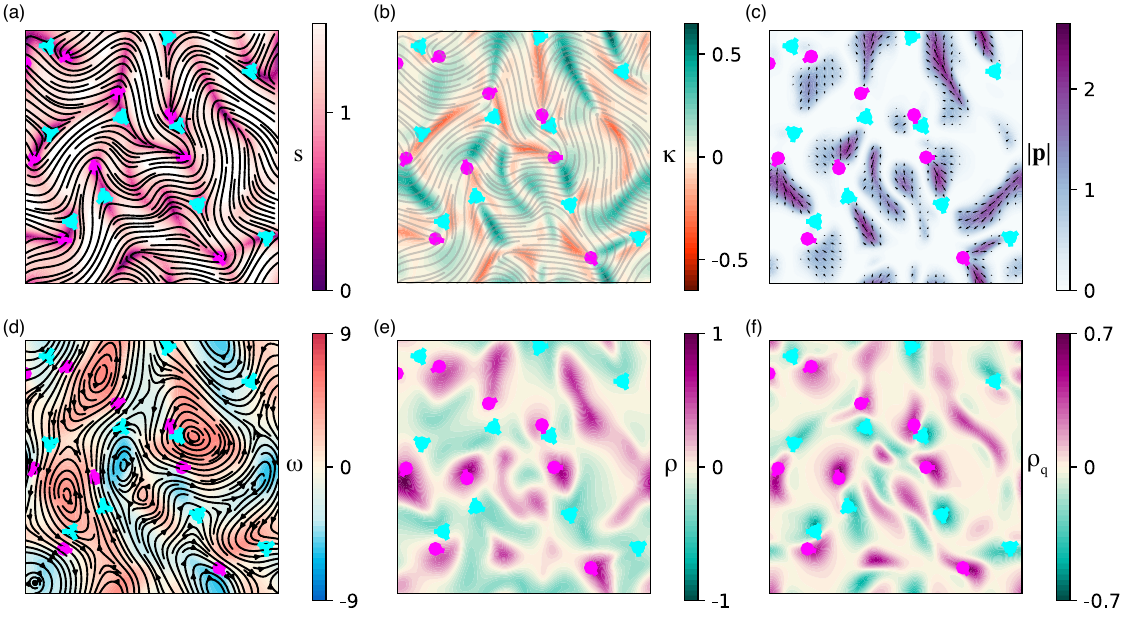}
\vspace{-9mm}
\caption{(a) Snapshot of nematic orientation $\bQ$ and degree of order $s$; (b) bend-splay field $\kappa$ (\eqn\ref{eqn:kappa}) where magenta tones indicate splay-rich regions and green tones are bend-rich; (c) degree of polar order $|\bp|$ and orientation $\bp$; (d) active flow $\bu$ and vorticity $\omega$; (e) defect-sensing field $\rho$; and (f), diffuse charge density $q$ (\eqn\ref{eqn:defdensity}). $\pm\frac{1}{2}$ defects are labeled in all panels with, respectively, magenta arrows and cyan three-pronged glyphs. The parameters used in this simulation are $k_{\rho_-}=100$ and $D_{\rho}=100$; all other parameters remain fixed and are listed in Table \ref{table:params}. Dynamics are shown in movie S1 \cite{esi}}.
\label{fig:hydro}
\end{figure*}

\section{Model}

\subsection{Nematohydynamics}
Our defect-sensing system will act on a liquid crystal whose dynamics are governed by the  active nematohydrodynamic equations in the high Ericksen and low Reynolds number limits \cite{BerisEdwardsThermo,koch2021}. We briefly, restate the equations here. In these limits, inertia and liquid crystalline stresses are omitted from the force balance. Nematic order is represented by the traceless and symmetric second order tensor $\bQ=s\left[\bn\otimes\bn-\bI/2\right]$ where $\bn$ is the nematic orientation and $s=\sqrt{2\bQ:\bQ}$ is the degree of order. Momentum conservation and incompressibility govern the fluid flow $\bu$ and hydrostatic pressure $p$
\begin{align}
\partial_t\bQ+\bu\cdot\nabla\bQ+\left[\bOmega,\bq\right]-\lambda\bE&=\left(1-s^2\right)\bq+\nabla^2\bQ, \nonumber \\
\nabla^2\bu-\nabla p - \alpha\nabla\cdot\bq -\xi \bu&=0, \nabla\cdot\bu=0,
\end{align}
where $2\Omega_{ij}=\partial_{j}u_i-\partial_{i}u_j$ and $2 E_{ij}=\partial_{j}u_i+\partial_{i}u_j$ are, respectively, the anti-symmetric and symmetric parts of the flow field gradient, $\xi$ is the strength of substrate friction, $[\cdots]$ is the commutator, and $\lambda=1$ is the flow alignment parameter.

\subsection{Defect Sensing System}




The defect detection system is governed by two PDEs. The first describes the orientation and activation strength of the dipolar enzyme through the vector $\bp$. The second PDE is a reaction-diffusion equation describing the resulting distribution of the chemical signal $\rho$, which we term the defect-sensing field. In a biological context, $\rho$ represents a morphogen, which could, in turn, couple to other processes. Here, we focus on its production only and leave it uncoupled to downstream dynamics. All equations are presented in dimensionless form, where the nematic coherence length and nematic relaxation time have been used, respectively, as the length and time scales \cite{koch2021}.

To begin, we consider the conditions for polar order to emerge in our system. Specifically, we let polarization develop in nematic regions where the local bend deformation exceeds a defined threshold. We quantify bend deformations with the scalar 
\begin{equation}
\kappa=Q_{ij}Q_{ik,j}Q_{kl,l}.
\label{eqn:kappa}
\end{equation}
We note that for perfectly ordered nematics, the quantity $|\bn\times\left(\nabla\times\bn\right)|$ would be a suitable measure of bend \cite{DeGennes1995}. In our case, we need a measure in terms of $\bQ$ that does not diverge near defects. We choose the scalar invariant $\kappa$ because it is the lowest order term that breaks splay-bend degeneracy \cite{zhang2017, Copic2020}. Typically, this term combines with others to represent the total bending energy; we use it in isolation here as a minimal geometric selection criterion. We manually tune $\kappa_0$ to select regions of high bend distortion outside defect cores. The second column of \fig\ref{fig:isolated} plots $\kappa$ fields of idealized defects and the dotted contour highlights $\kappa_0$.

The bulk free energy of the polar system is given by $\mathscr{F}_{\text{p,bulk}}=E_{\text{p1}}\left(1/2\right)\left(1+\left(1/2\right)|\bp|^2-\kappa/\kappa_0\right)|\bp|^2$,
where $\kappa_0$ is the critical curvature at polar order is created, which represents the activation of the enzyme. We choose this form for the $\kappa$-dependence because it provides a switch-like nonlinear behavior. Physically, this bulk free energy captures the stress/strain-dependent state of our constituents.

To complete the description of $\bp$, we need to specify its orientation with respect to $\bQ$. Using the schematic in \fig\ref{fig:schematic}c as a guide, we see that rod-like particles naturally acquire a polar axis $\perp \bQ$ when bent. Taking inspiration from polarization that arises in flexoelectric and bent-core nematics, we enforce this relationship by introducing $\bp-\bQ$ coupling terms \cite{Longa2016, Pajak2018,Rudquist1994,Meyer1969}. We align the molecules perpendicular to the director field with the term $\propto Q_{ij}p_ip_j$ the parameter $E_{\text{pQ1}}>0$. The following term, $\propto Q_{ij,j}p_i$, breaks $\bp\perp\bn$ degeneracy (\emph{i.e.} the orientation of $\bp$ is not uniquely prescribed by $\bp\perp\bn$ alone, see \fig S1 \cite{esi} for details). With no loss in generality, we arbitrarily choose an electrostatic sign convention such that $\bp$ will point towards $-\frac{1}{2}$ defects and away from $+\frac{1}{2}$ defects by letting $E_{\text{pQ2}}>0$ in our simulations. At the molecular scale, this broken symmetry could arise from a one-way hinge such that, in our schematic, the cyan side of the enzyme always resides on the exterior and magenta, on the interior. The complete free energy for the polar field is then
\begin{multline}
\mathscr{F}_{\text{p}}=E_{\text{p1}}\frac{1}{2}\left(\frac{1}{2}|\bp|^2+1-\frac{\kappa}{\kappa_0}\right)|\bp|^2+ E_{\text{p2}}\frac{1}{2}|\nabla\bp|\\
+E_{\text{pQ1}}\frac{1}{2}\left(\bp\otimes\bp\right):\bQ
+E_{\text{pQ2}}\left(\nabla\cdot\bQ\right)\cdot\bp
.
\label{eqn:polarenergy}
\end{multline}
 The dynamics of $\bp$ are given by the combination of gradient descent and convective dynamics 
\begin{multline}
\partial_t\bp+\bu\cdot\nabla\bp+\bOmega\bp=-\frac{1}{\gamma}\frac{\delta\mathscr{F}_{\text{p}}}{\delta \bp}=
\beta_{\text{p1}}\left(\frac{\kappa}{\kappa_0}-1-|\bp|^2\right)\bp
\\+\beta_{\text{p2}}\nabla^2\bp
-\beta_{\text{pQ1}}\bQ\bp
-\beta_{\text{pQ2}}\nabla\cdot\bQ,
\end{multline}
where coefficients $\beta_{\cdots}=\gamma^{-1}E_{\cdots}$. The third column of \fig\ref{fig:isolated} shows the steady state magnitude (color field) and direction (vectors) of $\bp$ around stationary $\pm\frac{1}{2}$ defects. For simplicity, we assume that the emergence of polar order does not apply torque to the nematic such that the dynamics of $\bQ$ and $\bu$ are independent of $\bp$.

Finally, we consider the chemical activity of the oriented constituents. We let $\bp$ represent the strength and orientation of an active reaction dipole. In other words, each element consists of an infinitesimal source-sink pair where each pole of the element either catalyzes the production (head of dipole) or destruction (tail) of a species $a$. We additionally include a background process that maintains a concentration $a_0$, so we focus on the dynamics of the deviation described by the order parameter $\rho= (a-a_0)/a_0$. Analogous to polarization charge in electrostatics, no net production or depletion of $\rho$ occurs where the dipole distribution $\bp$ is uniform in both strength and orientation. However, where distortions in orientation or degree of polarization exist, the reaction is biased depending on the local orientation of the molecules. Formally, the polarization charge density-like term $\nabla\cdot\bp$ drives changes in $\rho$. The full transport equation for $\rho$ is given by
\begin{align}
\partial_t\rho+\bu\cdot\nabla\rho&=k_{\rho^+}\nabla\cdot\bp-k_{\rho^-}\rho+
D_{\rho}\nabla^2\rho,
\label{eqn:rhodynamics}
\end{align}
where the reaction $k_{\rho^-}$ scales the background process that drives $\rho\rightarrow 0$ ($a\rightarrow a_0$) and $D_{\rho}$ is Fickian diffusion.

We note a mathematical similarity between our proposed system and work on living active nematics \cite{Genkin2017}. There, the transport of bacteria embedded in a passive liquid crystal gave rise to preferential accumulation and depletion of bacteria in, respectively, $+\frac{1}{2}$ and $-\frac{1}{2}$ defects. Importantly, accumulation is related directly to the motility and active stress in the sample. In contrast, our model decouples the driver of concentration gradients from the source of hydrodynamic activity. Related models of active nematohydrodynamics that consider variable density, to our knowledge, do not produce density fields that correlate strongly with defect type \cite{Thampi2015,Putzig2015}. Particles embedded in passive \cite{wang2016c} and active nematics \cite{foffano2019} can label defects by aggregating in regions of disorder; however the processes that highlight defects operate on different principles than the biological examples that inspire this work.


\begin{figure*}
\includegraphics[width=\textwidth]{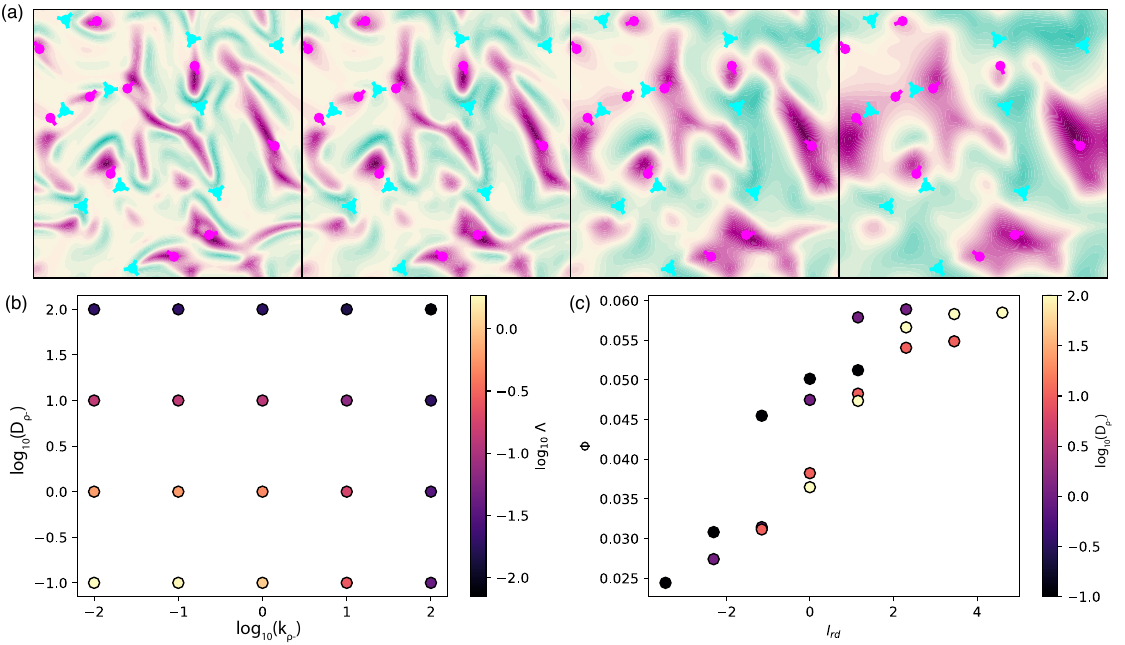}
\vspace{-9mm}
\caption{(a) snapshot of normalized $\rho$ for varying diffusion rate $D_{\rho}=\{0.1,1,10,100\}$ (from left to right) for fixed degradation rate $k_{\rho^-}=10$, (b) maximum amplitude of $\rho$, $\Lambda=\max_t\left(\max_\mathbf{x}{\rho}-\min_\mathbf{x}{\rho}\right)/2$, and (c) error metric $\Phi$ as a function of reaction-diffusion screening length $l_{\text{RD}}=\sqrt{D_{\rho}/k_{\rho^-}}$, where the color bar indicates $D_{\rho}$.}
\label{fig:RDphase}
\end{figure*}

\section{Results}
To demonstrate the behavior of the system, we first consider its steady state behavior in response to idealized, static defects. We prescribe the nematic order tensor to be $\bQ=s\left(\bx\right)\left[\bn\otimes\bn-\frac{1}{2}\bI\right]$, where the defect core is approximated using a Gaussian divot with unit standard deviation such that $s\left(\bx\right)=1-e^{-\frac{1}{2}\bx\cdot\bx}$. The director fields are given by the analytical expressions $\bn_{\pm}=\{-\cos\left(\frac{1}{2}\left(\theta-\theta_0\right)\right),\mp\sin\left(\frac{1}{2}\left(\theta-\theta_0\right)\right) \}$ with $\theta=\tan^{-1}\left(y/x\right)$ and orientation $\theta_0 = \pm\pi/2$ for $\pm\frac{1}{2}$. 

\fig\ref{fig:isolated}a shows these idealized defects and the responding fields. All simulations are performed using the open source spectral-based solver Dedalus \cite{Burns2020}.  The red circle throughout the panels indicates the defect core size by plotting the $s=0.75$ level-set. \fig\ref{fig:isolated}b shows contours of $\kappa$ with bend-rich regions in teal, and splay-rich in orange. The dotted line highlights the chosen critical value $\kappa_0$, above which polar order emerges. Note that the $+\frac{1}{2}$ defect possess one large bend-dominated region while the $-\frac{1}{2}$ possess three such regions of smaller size; we will discuss the consequences of this difference shortly. Next, \fig\ref{fig:isolated}c shows the degree $|\bp|$ (color field) and orientation of polarization $\bp$ (vector field). The direction of polar order follows the sign convention we chose above, with $\bp$ pointing away from $+\frac{1}{2}$ defects and towards $-\frac{1}{2}$ defects. The final column of \fig\ref{fig:isolated} shows the distribution of $\rho$. Purple contours highlight dipole-rich regions by plotting the $|\bp|=1$ level-set. Cyan and magenta contours, $\rho^*=\pm0.65$, emphasize the asymmetry of the resulting concentration field around each defect: $|\rho|>|\rho^*|$ inside defects core and $|\rho|<|\rho^*|$, outside. 

\begin{figure}
\includegraphics[width=\columnwidth]{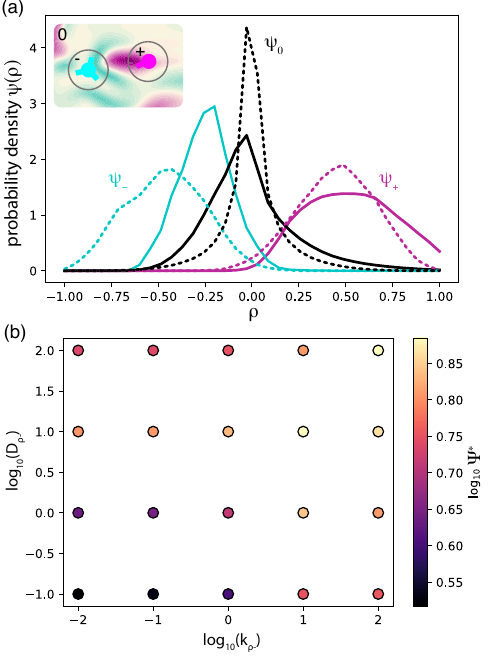}
\vspace{-9mm}
\caption{(a) Probability density functions $\psi\left(\rho\right)$ in regions containing $-\frac{1}{2}$ defects (dark cyan), $+\frac{1}{2}$ defects (purple), or no defects (black) quantifies the performance of the defect detection system (solid) in the reaction-diffusion-dominated regime ($D_{\rho}=100$ and $k_{\rho^-}=100$) compared to diffuse charge density $\rho_q$ (dashed). Inset: regions are considered to belong to a defect if they fall within one nematic coherence length of a defect core ($l_c=1$ in our simulations); (b) metric quantifying distinctness between regions $\Psi^*$; larger values indicate less overlap between distributions and, therefore, strong defect-identification ability. The best performance occurs in the reaction-diffusion-dominated regime.}
\label{fig:histogram}
\end{figure}


We now consider the dynamic scenario. The hydrodynamic flows created by activity result in continuous creation and annihilation of $\pm\frac{1}{2}$ defect pairs; \fig\ref{fig:hydro}a,d show typical nematic order and flow fields, see also supplemental movies S1 and S2 \cite{esi}. Defects rarely possess geometry identical to their idealized forms considered in \fig\ref{fig:isolated}. Consequently, regions of high polar order $|\bp|$ take on large-scale, extended forms that populate the space between defects, \fig\ref{fig:hydro}c. Despite this, these regions remain approximately polar themselves with poles residing near defects that serve as local sources and sinks for $\rho$, which is sufficient for the system to report on the local topological environment, \fig\ref{fig:hydro}e.


We note that inherent geometric differences between defect types give rise to different $\rho$ signal strengths. The three smaller regions of $|\bp|$ near $-\frac{1}{2}$ defects, even in the idealized case, \fig\ref{fig:isolated} have a larger surface area to volume ratio that leads to greater diffusive loss and weaker $\rho$ signal compared to $+\frac{1}{2}$ defects. This becomes more pronounced when hydrodynamics are present, which further distort the bend-rich sources of $\rho$.

As a point of comparison, in \fig\ref{fig:hydro}f, we plot a local measure of the defect charge density proposed in Ref. \cite{Blow2014}
\begin{equation}
\rho_q=\frac{1}{4\pi}\left(\partial_{x}Q_{x\alpha}\partial_{y}Q_{y\alpha} - \partial_{x}Q_{y\alpha}\partial_{y}Q_{x\alpha}\right).
\label{eqn:defdensity}
\end{equation}
We compare $\rho$ to this field because the latter follows directly from the definition of the winding number and because both $\rho_q$ and $\rho$ utilize similar levels of information on the director (i.e. both use $\nabla\bQ$). We expect $\rho_q$ to serve as an approximate upper bound on performance because it depends only on $\bQ$ instantaneously and is not subject to rates associated with mass transport. Indeed, we find a qualitative correspondence between $\rho_q$ and $\rho$ in \fig\ref{fig:hydro}e,f. Both fields report on the charginess of a region, and often exhibit extrema shortly before nucleation events and after annealing.

The performance of the system naturally depends on the various parameters in the model. In \fig\ref{fig:RDphase}, we vary the degradation rate $k_{\rho^-}$, and diffusion coefficient $D_{\rho}$ and readily observe a change in $\rho$'s spatial distribution (\fig\ref{fig:RDphase}a) and amplitude (\fig\ref{fig:RDphase}b). The motivation for focusing on these two parameters is discussed in Appendix \ref{sec:appendix_parameters}. Briefly,  we consider the limiting case where $\bp$ dynamics are approximately locked to $\bQ$ and, because $k_{\rho^+}$ can be removed by rescaling $\rho^*=\rho/k_{\rho^+}$, set $k_{\rho^+}=1$ with no loss in generality. The latter motivates our focus on comparing normalized quantities; the scale of $\rho$ can always be trivially modified by changing $k_{\rho^+}=1$ without impacting its spatial structure.

We quantitatively compare $\rho$ and $\rho_q$ by defining the metric $\Phi=\langle\left( \bar{\rho}-\bar{\rho}_q\right)^2 \rangle$, where the overbar denotes quantities normalized by the maximum amplitude (\emph{e.g.} $\bar{\rho}=\rho / \Lambda$, where $\Lambda=\max_t\left(\max_\mathbf{x}{\rho}-\min_\mathbf{x}{\rho}\right)/2$). Fig.\ref{fig:RDphase} shows that the trend in $\Phi$ is partially collapsed by plotting it as a function of the reaction-diffusion decay length scale $l_{\text{rd}}=\sqrt{D_{\rho}/k_{\rho^-}}$. For this particular error metric, the trend shows that shorter reaction-diffusion lengths favor agreement between $\rho$ and $\rho_q$. This is reasonable because larger $l_{rd}$ destroy the fine structure of $\rho$ that is present in $\rho_q$. The first few panels of \ref{fig:RDphase}a, for example, shows the emergence of fine structure in $\rho$; additional snapshots from the $\{k_{\rho^-},D_{\rho}\}$ phase space are shown in supplemental \fig S2. However, the balance of reaction and diffusion processes cannot alone determine the dynamics of quality of $\rho$ because of convection. Furthermore, the system-wide error metric doesn't capture the relationship between $\rho$ and the true defect locations.

To refine our assessment of $\rho$, we examine its values in the neighborhood of defects by building probability distributions $\psi\left(\rho\right)$ within circles of unit radius (one nematic coherence length) around each defect type and compare those to the distributions of the region outside those circles (inset of \fig\ref{fig:histogram}a). \fig\ref{fig:histogram}a shows the probability distributions $\psi_+,\psi_-,\psi_0$ where subscripts, respectively, denote $+\frac{1}{2}$, $-\frac{1}{2}$ half, and neutral regions. We find that the distributions (solid) of defect-containing regions possess peaks at positive and negative $\rho$ values and that $\rho=0$, on average, in neutral regions. Ideally, these distributions would not overlap at all as this would imply that a given $\rho$ value could always be associated with either a defect or neutral region. Instead, the distributions overlap, so there is some ambiguity. For comparison, we also plot the distributions for $\rho_q$ (dashed lines in \fig\ref{fig:histogram}), which also overlap but to a lesser degree. 

We are primarily interested in measuring the overlap of these distributions: the less they overlap, the better the defect-sensing system is at discriminating between $\pm\frac{1}{2}$ defect-containing and defect-free regions. To quantify, we introduce the metric $\Psi=\psi_{+,-}+\psi_{+,0}+\psi_{-,0}$ where $\psi_{ij}=|\psi_i-\psi_j|$. We use this particular measure because it is maximized when the distributions are disjoint but agnostic to any further separation. We plot the normalized value $\Psi^*=\Psi/\Psi_q$ (where $\Psi_q$ is the value obtained for the diffuse charge density $\rho_q$, \eqn\ref{eqn:defdensity}) as a function of degradation rate and diffusion coefficient, \fig\ref{fig:histogram}b. We readily see that the best differentiation between regions occurs in the upper-right of the figure, where we expect reaction and diffusion  transport to dominate over convection. For the average flow rates we observe ($u\sim\mathcal{O}\left(1\right)$) we indeed find that the Peclet number $\text{Pe}\sim u l / D_{\rho} \sim \mathcal{O}\left(10^{-1}\right)$ and the inverse Damk{\"o}hler number $\text{Da}^{-1}\sim u/l k_{\rho^-}\sim O(10^{-1})$ are both small over a length scale of ten defect cores $l \sim \mathcal{O}\left(10\right)$ indicating that, respectively, diffusion and reaction dynamics dominate for the highest values we consider. 

We conclude that in the RD-limited regime, our proposed active reaction-diffusion model performs nearly as well as a classical measure of topological charge. To close, we speculate that increasing performance further likely requires additional physics. In Appendix \ref{sec:appendix_chempotential}, we briefly demonstrate that augmenting $\rho$'s dynamics with an attractive potential that induces phase separation enhances localization of $\rho$. Thus, the design space for sensing systems is rich, encompassing both active elements and passive elements.




\section{Discussion and Conclusion}

Steering the dynamics of active materials requires strategies distinct from those for near-equilibrium systems. Experimental and computational works have demonstrated that spatiotemporally actuating local force generation offers new avenues for both ad hoc \cite{Zhang2019, ross2019, lemma2023, zarei2023} and computationally-derived control schemes \cite{Norton2020, falk2021a, Shankar2022, shankar2024, ghosh2024}. However, these studies have relied on externally sensed global flow fields or director information to devise their respective, exogenously implemented control strategies. In order to create truly autonomous biopolymeric materials \cite{zhang2017, Lemma2021, lemmabez2022, Lee2021, Hsu2022, Berezney2022}, the control must be implemented endogenously, only relying on internal sensing.

By demonstrating that a material's topological state can be sensed internally, we posit that the incorporation of systems biology principles into active materials becomes more accessible. The biochemical basis of many fundamental functional motifs, such as oscillators and switches, is known \cite{Tyson2003a}. One can envision chaining together such elements with $\rho$ to create spatiotemporally complex dynamics that, in turn, pattern the active stress strength $\alpha$ in novel ways. The design space of such systems is large and challenging to navigate. Tools from nonlinear dynamics and controls such as the theory of passivity and port-Hamiltonian systems may provide additional tools for rationally designing and analyzing closed-loop PDE systems \cite{Jovanovic2008, Ahmadi2016}. Additionally, knowledge of high-dimensional phase space of uncontrolled nematohydrodynamics may be leveraged to design embedded controllers that stabilize an unstable flow state or create pathways that connect existing states \cite{Norton2020, Wagner2021, Wagner2023}.

We motivated this study with examples of morphogenesis in plants and animals that demonstrate coupling between polar or nematic order and biochemical dynamics. The theme of coupling between form and chemistry is a biologically important task at the sub-cellular scale as well. Proteins whose binding or state depends on curvature are utilized extensively in living systems (\emph{e.g.} septins \cite{shi2023, nakazawa2023}, BAR \cite{Jin2022}, and PIEZO 1/2 \cite{qin2021}). 
In Appendix \ref{sec:appendix_binding}, we briefly consider a modified model that explicitly includes binding/unbinding dynamics of the flux-producing molecules to the nematic. The ubiquity of proteins with curvature-dependent properties highlights the potential for building active systems with direct feedback from geometry and topology. Further, these examples can serve as starting points for the design of \emph{de novo} proteins \cite{pillai2023} that fulfill functional roles in synthetic materials not provided by nature. 

We close by highlighting that our proposed sensing system is, like the embedding nematic, active.  The species $\rho$ is continuously produced and destroyed at the poles of the dipoles (\eqn\ref{eqn:rhodynamics}), or in a variation, pumped against a gradient (Appendix \ref{sec:appendix_flux}). Consequently, there is an energetic cost to operating this system. Understanding how biological reaction-diffusion systems use energy to sense external stimuli \cite{hathcock2023a} or internal geometry \cite{ramm2019} is an active area of research. We hope that our work inspires further investigations into non-equilibrium geometric feedback systems and their potential to shape active material dynamics.


\begin{acknowledgments}
We thank the participants and organizers, Linda Hirst and Kevin Mitchell, of the Telluride 2021 Workshop \emph{Nonlinear Dynamics of Active Matter} for stimulating discussions that lead to these ideas. We thank Brian Camley and Philip Nelson for helpful conversations. This material is based upon work supported by the U.S. Department of Energy, Office of Science, Office of Basic Energy Sciences under Award No. DE-SC0022280.
\end{acknowledgments}

\bibliographystyle{ieeetr}

\appendix

\section{Parameter Selection\label{sec:appendix_parameters}}

The model we have developed contains eight free parameters,$\{\kappa_0,\beta_{\text{p}1},\beta_{\text{p}2},\beta_{\text{pQ}1},\beta_{\text{pQ}2},k_{\rho^+},k_{\rho^-},D_{\rho}\}$, that need to be fixed or explored. Further, the non-dimensional nematohydrodynamic equations possess three parameters $\{\lambda,\alpha,\xi\}$. Table \ref{table:params} summarizes our parameter choices; we now outline our process for determining their values. Because the behavior of the active nematic is not the focus of this study, we keep the geometry produced by the nematic ``fixed'' by picking values for the friction $\xi$, activity $\alpha$, flow alignment $\lambda$. We then tune the value of $\kappa_0$ such that we have localized regions of bend above this critical value near both kinds of defects. 

Next, we restrict our choices of parameters governing $\bp$ by working towards certain limits.  Because we envision $\bp$ as being composed of elements completely embedded in the nematic, we assign $\beta_{\text{p}1,2}=1$ such that the reorientation  and bulk dynamics of $\bp$ are comparable with $\bQ$. Next, to match our physical picture and ensure that $\bq\perp\bn$ (the axis of bend is always perpendicular to the alignment of the rod, \fig\ref{fig:schematic}), we let $\beta_{pQ1}\gg1$. Finally, we choose the alignment of $\bp$ with the divergence of $\bQ$ to be strong, but not so strong as to override the perpendicular requirement: $\beta_{pQ1}>\beta_{pQ2}>1$. See \fig S1 for additional details \cite{esi}.

Finally, we consider the three parameters governing the dynamics of $\rho$ in \eqn\ref{eqn:rhodynamics}. We eliminate one, $k_{\rho^+}$, by noting that it provides a natural scale for $\rho$. In the paper, since we examine normalized $\rho$ when comparing simulation results, we arbitrarily set $k_{\rho^+}=1$ without any loss in generality. This leaves only the coefficients $D_{\rho},k_{\rho^-}$ as free variables.  We vary the coefficients for each over a few orders of magnitude to explore the performance of the defect-sensing system in various regimes. 

\begin{table}[h]
\begin{center}
\begin{tabular}{ |c|c|c| } 
 \hline 
parameter description & symbol & value \\
 \hline 
 active stress strength & $\alpha$ & -6 \\
 substrate friction & $\xi$ & 0.1 \\
 flow alignment & $\lambda$ & 1 \\
 \hline 
 critical bend & $\kappa_0$ & 0.1\\
bulk polarization relaxation rate & $\beta_{\text{p1}}$ & 1\\
orientational diffusivity of $\bp$ & $\beta_{\text{p2}}$ & 1\\
alignment rate of $\bp$ to $\bQ$ & $\beta_{\text{pQ1}}$& 50 \\
alignment rate of $\bp$ to $\nabla\cdot\bQ$ & $\beta_{\text{pQ2}}$ & 5\\
 \hline 
$\rho$ production rate & $k_{\rho^+}$ & 1 \\
$\rho$ degradation rate & $k_{\rho^-}$ & 0-100\\
$\rho$ diffusivity & $D_{\rho}$ & 0.01-100 \\
\hline
\end{tabular}
\end{center}
\caption{Table of parameters used in simulations grouped by their appearance in the (top) nematohydrodynamic equations, (middle) $\bq$-equation, and (bottom) $\rho$-equation; all parameters are fixed throughout the main paper except for the final two: $k_{\rho^-}$ and $D_{\rho}$. \label{table:params}}
\end{table}

\section{Model Variations\label{sec:appendix_variations}}

We briefly introduce three variations of the model with qualitatively similar dynamics that correspond to different physical realizations of the system (Appendices \ref{sec:appendix_flux} and \ref{sec:appendix_binding}) or add additional physical processes that improve performance (Appendix \ref{sec:appendix_chempotential}). 

\subsection{Active flux \label{sec:appendix_flux}}
For this case, we consider $\bp$ to represent a vector along which active transport of a species $\bar{\rho}$ occurs, rather than a density of source-sink dipoles. The the total flux of $\bar{\rho}$ is given by
\begin{equation}
\mathbf{j}_{\rho}=\bar{\rho}\left(\bu+\chi\bp\right)-D_{\rho}\nabla \bar{\rho}.
\end{equation}
We let $\bar{\rho}$ be uniformly generated and consumed so as to be driven towards a steady state concentration $\rho_0$ at rate $k_{\rho}$. The total dynamics for $\rho$ are then.
\begin{equation}
\partial_t\bar{\rho} = -\nabla\cdot \mathbf{j}_\rho + k_{\rho} \left(\rho_0 -\bar{\rho}\right).
\end{equation}
To focus on topology-induced behavior, we define a new variable $\rho=\bar{\rho}-\rho_0$. Substituting this change of variables and the flux equation into the conservation equation gives

\begin{equation}
\partial_t\rho +\left(\bu +\chi\bp\right)\cdot\nabla\rho = -\chi\rho_0\nabla\cdot\bp - k_{\rho} \rho
+D_\rho \nabla^2 \rho.
\end{equation}
Importantly, this transport equation contains nearly identical non-equilibrium terms as the main model; with concentration and dilution of $\rho$ governed by the divergence of the field $\bp$. An example of the dynamics with $\chi=1$ and $\rho_0=1$ is shown in the supplemental movie S3 \cite{esi}.

\subsection{Curvature-dependent binding\label{sec:appendix_binding}}
In the main body, we considered the active dipole to be activated directly by bend deformations. Taking inspiration from curvature-dependent assembly on membranes, we consider a situation where the active dipoles are an additional constituent, a species of curved molecules that preferentially bind to regions of the nematic matching their intrinsic curvature. Dynamics are given by the transport equations
\begin{align}
\partial_t\phi+\bu\cdot\nabla\phi &= D_{\phi}\nabla^2\phi \label{eq:phidynamics}\\
&+ k_{\phi^+}\left(1-\phi\right)f\left(\kappa\right) - k_{\phi^-}\phi\nonumber,\\
f\left(\kappa\right) &= e^{-\left(\kappa-\kappa_0\right)^2/\delta},\label{eq:phion}
\end{align}
where the on-rate is limited by the carrying capacity of the surface set by the factor $\left(1-\phi\right)$ and geometric specificity is captured through a geometry-dependent on rate that is maximal at $\kappa=\kappa_0$; where $\delta$ defines the degree of specificity and depends the elasticity of the curved particles and embedding nematic.

\begin{figure}
\includegraphics[width=\columnwidth]{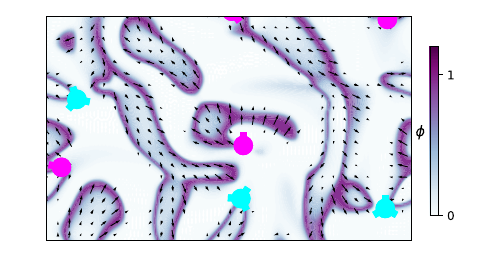}
\vspace{-9mm}
\caption{Concentration of bound molecules (color field) and polar order (arrows) for the model variation described be \eqns\ref{eq:phidynamics}-\ref{eq:phipol}. The resulting dynamics of the model are shown in the supplemental movie S4 \cite{esi}}.
\label{fig:curv-dependent}
\end{figure}

To complete this version of the model, we modify the free energy of the polar field by replacing the explicit $\kappa$-dependence with a lyotropic bulk free energy term that builds polar order where $\phi$ surpasses a critical concentration $\phi_c$
\begin{equation}
\mathscr{F}_{\text{p,bulk}}=E_{\text{p1}}\frac{1}{2}\left(\frac{1}{2}|\bp|^2+1-\frac{\phi}{\phi_c}\right)|\bp|^2.
\label{eq:phipol}
\end{equation}
\begin{multline}
\partial_t\bp+\bu\cdot\nabla\bp+\bOmega\bp=-\frac{1}{\gamma}\frac{\delta\mathscr{F}_{\text{p}}}{\delta \bp}=
\beta_{\text{p1}}\left(\frac{\phi}{\phi_c}-1-|\bp|^2\right)\bp
\\+\beta_{\text{p2}}\nabla^2\bp
-\beta_{\text{pQ1}}\bQ\bp
-\beta_{\text{pQ2}}\nabla\cdot\bQ,
\label{eq:phipol2}
\end{multline}
The remaining dynamics are unchanged from the model presented in the main body. \fig\ref{fig:curv-dependent} shows a snapshot of the $\phi$ and $\bp$ fields for this model variation. In contrast to \ref{fig:hydro}c, $\bp$ is isolated to regions where $\kappa\sim\kappa_0$. Despite this, $\rho$ performance remains qualitatively similar to that of the primary model considered. An example of the dynamics is shown in supplemental movie S5 \cite{esi}.


\subsection{Increasing $\rho$ signal contrast \label{sec:appendix_chempotential}}

\begin{figure}
\includegraphics[width=\columnwidth]{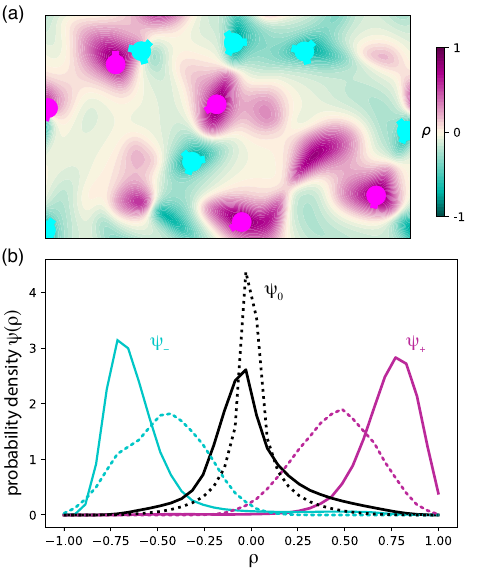}
\vspace{-9mm}
\caption{Comparison of distributions $\psi_{+,-,0}$ between $\rho$ with phase separation dynamics (solid) and diffuse charge density $\rho_q$ (dashed) shows that phase separation enhances performance ($\psi_{+,-,0}$ becomes more disjoint). Dynamics are shown in supplemental movie S5 \cite{esi}.}
\label{fig:chempotential}
\end{figure}

In Figs.\ref{fig:RDphase} and \ref{fig:histogram}, we varied the reaction rates and diffusivity of $\rho$, but were not able to approach the same level of discrimination between regions containing defects as the diffuse-charge density. To experiment with bioinspired methods of enhancing contrast, here, we add attractive interactions to the signaling molecules $\rho$ in the form of a Cahn-Hilliard-like potential that is present only in regions of nematic disorder (\emph{i.e.} $s<s_c$). This modifies the dynamics of $\rho$ in the following way

\begin{align}
\partial_t\rho+\bu\cdot\nabla\rho&=k_{\rho^+}\nabla\cdot\bp-k_{\rho^-}\rho+
D_{\rho}\nabla^2\mu_{\rho},\\
\mu_{\rho}&=\gamma_{\rho}\nabla^2\rho+\left(s-s_c\right)\rho+A\rho^2+\rho^3,
\end{align}
where $\mu_{\rho}$ is the chemical potential of $\rho$. In \fig\ref{fig:chempotential}a, we show the resulting $\rho$ field whose maxima are noticeably more localized to the defects than the version presented in \fig\ref{fig:hydro} (see supplemental movie S4 for dynamics). Further, \fig\ref{fig:chempotential}b shows the probability density function of $\rho$ compared to the case when $\rho$ is purely diffusive. We see a substantial decrease in the overlap between the distributions. In fact $\Psi^*=1.08$, indicating that the system identifies regions with defects better than the diffuse charge density. This suggests that when designing biomimetic sensing system that, cooperative interactions can improve performance. 

\bibliographystyle{ieeetr}


\end{document}